\documentclass{osa-article}

\journal{osac}
\articletype{Research Article}

\begin{document}

\title{Design and analysis of guided modes in photonic waveguides using optical neural network}

\author{Nusrat Jahan Anika,\authormark{1,*} and Md Borhan Mia,\authormark{2}}

\address{\authormark{1}Department of Software Engineering, Daffodil International University, Dhaka 1205, Bangladesh\\
\authormark{2}Department of Electrical and Computer Engineering, Texas Tech University, Lubbock, Texas 79409, USA}
\email{\authormark{*}nusrat35-1977@diu.edu.bd} 

\begin{abstract}
We present a deep learning approach using an optical neural network to predict the fundamental modal indices $n_{\rm{eff}}$ in a silicon (Si) channel waveguide. We use three inputs, e.g., two geometric and one material property, and predict the $n_{\rm{eff}}$ for transverse electric and transverse magnetic polarizations. With the least number (i.e., $3^3$ or $4^3$)  of exact mode solutions from Maxwell's equations, we can uncover the solutions which correspond to $10^3$ numerical simulations. Note that this consumes the lowest amount of computational resources. The mean squared errors of the exact and the predicted results are $<10^{-5}$. Moreover, our parameters' ranges are compatible with current photolithography and complementary metal-oxide-semiconductor (CMOS) fabrication technology. 
We also show the impacts of different transfer functions and neural network layouts on the model’s performance.
Our approach presents a unique advantage to uncover the guided modes in any photonic waveguides within the least possible numerical simulations. 
\end{abstract}
\section{Introduction}
Deep learning (DL) \cite{lecun2015deep} is a class of techniques in machine learning. Due to effectiveness, adaptability, and computational speed, it has a wide range of applications in image and speech recognition \cite{hinton2012deep,krizhevsky2012imagenet,tompson2014joint}, decision-making \cite{silver2016mastering}, analyzing particle accelerator data \cite{adam2015higgs}, and predicting the activity of drug molecules \cite{ma2015deep}.
DL technique recently penetrates the other fields: biology \cite{ching2018opportunities}, material science \cite{ziletti2018insightful}, chemistry \cite{mater2019deep}, microscopy \cite{rivenson2017deep}, and photonics \cite{zibar2017machine,pilozzi2018machine}. More specifically, in optical communication, DL is proven to be invaluable for optoelectronic components characterizations, predicting performance, and optimizing. It is also applied in inverse design (e.i., multilayer structures) \cite{liu2018training}, waveguide geometry \cite{alagappan2019deep}, and metamaterial structures \cite{ma2018deep}. To design these devices, we perform 3D finite-difference time-domain (FDTD) simulation to obtain the most accurate data. 
But it often takes a lot of time. Thus, we need to switch 2D FDTD, which is not always as good as 3D FDTD. Thus, we need to adopt a technique that provides the results within the least possible time. In this sense, the DL technique is a good candidate for accurate data predictions. 
Our article shows how the DL technique is applied on a photonic waveguide to
predict the modal indices with the least number of simulations.

\begin{figure}[ht]
\centering
\includegraphics[width=1\linewidth]{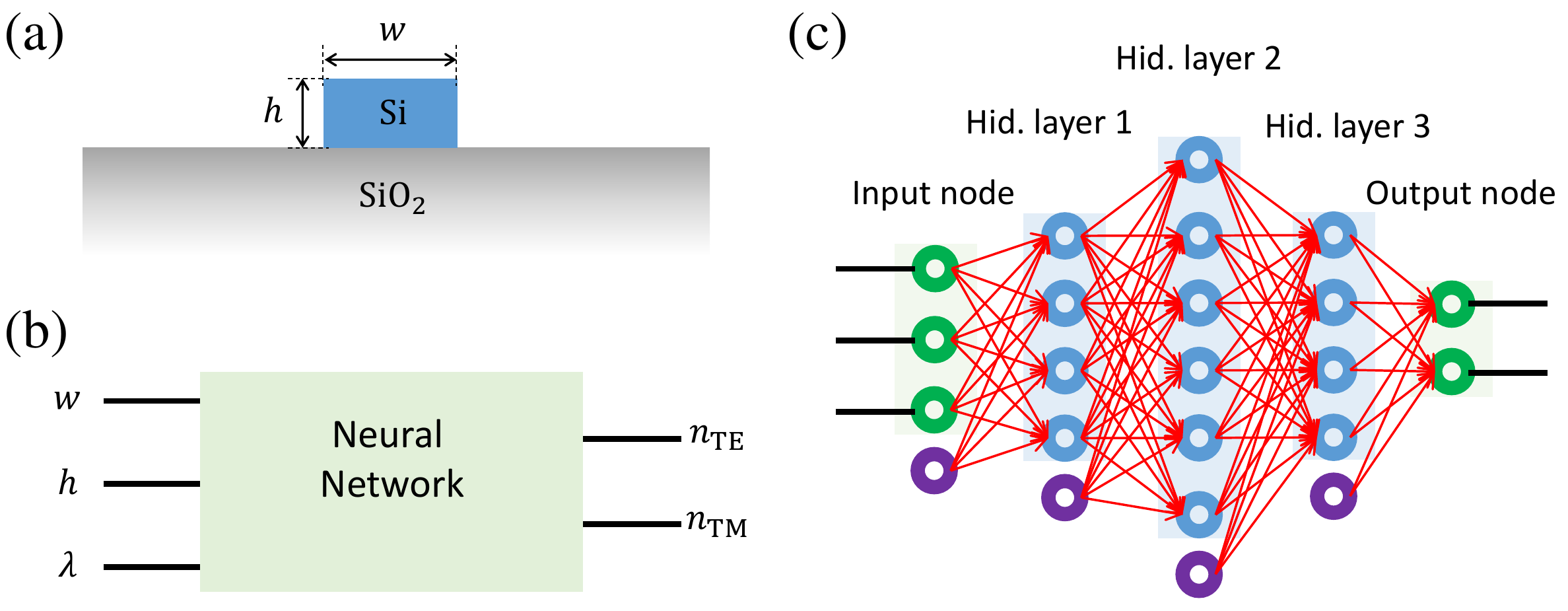}
\caption{\textbf{Waveguide schematic and Neural Network (NN) configurations}.
(a) The schematic cross-section of Si strip waveguide with geometric parameters are width $w$ and height $h$. (b) Proposed deep learning model to predict the fundamental effective mode indices for the transverse electric (TE) and transverse magnetic (TM) polarization of light, i.e., ${\it n}_{\rm{TE}}$ and $\it {n}_{\rm{TM}}$, respectively. Note that we use three inputs: width $w$, height $h$, and wavelength $\lambda$ for NN. (c) The suggested layout of the NN with three hidden layers, i.e., Hid. layer$=1,2,3$. Notice that blue circles indicate the neurons in each hidden layer, and green circles are for the input and output nodes. The purple circles in the input and hidden layers indicate bias, a constant term.}
\label{fig:NN-network}
\end{figure}

The basic structure of a dielectric waveguide consists of a core (e.g., high index medium) and a cladding (e.g., low index medium). A guided wave propagates in the longitudinal direction resulting in waveguide modes. The waveguide modes are the field patterns whose amplitude and polarization remain the same along the propagation direction. These field patterns can have different names based on their characteristics: transverse electric (TE) mode, transverse magnetic (TM) mode, transverse electric and magnetic (TEM) mode, and hybrid modes. In the channel waveguides (Fig.~\ref{fig:NN-network}(a)), there exist TE and TM modes only. These modes are solved using time-independent Maxwell's equations. Note that 1D slab waveguide modes can be easily solved using analytical solutions \cite{okamoto2006fundamentals}. However, for the 2D waveguides (i.e., channel waveguides), numerical simulation is a must for the exact solutions. More specifically, we use finite-difference \cite{yu2004mesh} and finite elements \cite{mabaya1981finite} methods to solve the waveguide modes numerically. These are well-established methods, often requiring too much computational time to optimize and sweep the geometric parameters. Moreover, we need to consider the fabrication limitations to design the parameters of a photonic waveguide. Hence, for a well-defined geometry, we need to perform a lot of numerical simulations. Many articles DL method using neural network (NN) techniques in predicting $n_{\rm{eff}}$, propagation loss, bending loss, and crosstalks in plasmonic and photonic waveguides \cite{gabr2019design,hammond2019designing,andrawis2016artificial}. 
However, there is no detail of how NN layout, hidden layers, and transfer functions can affect the prediction of NNs. 

In this paper, we explore a deep learning technique using an optical Neural Network (NN) to predict the fundamental effective indices for transverse electric (TE) and transverse magnetic (TM) modes of a photonic waveguide, with the least number of numerical simulations. 
We also show how different layouts of hidden layers and transfer functions affect the results of NN. 
Our DL technique with NN outperforms the conventional interpolation techniques while producing the lowest mean squared errors.
Moreover, the same technique could be useful to predict the other parameters, i.e., group index, group velocity, dispersions, and confinement/bending loss in waveguides.
\begin{figure}[hb]
\centering
\includegraphics[width=1\linewidth]{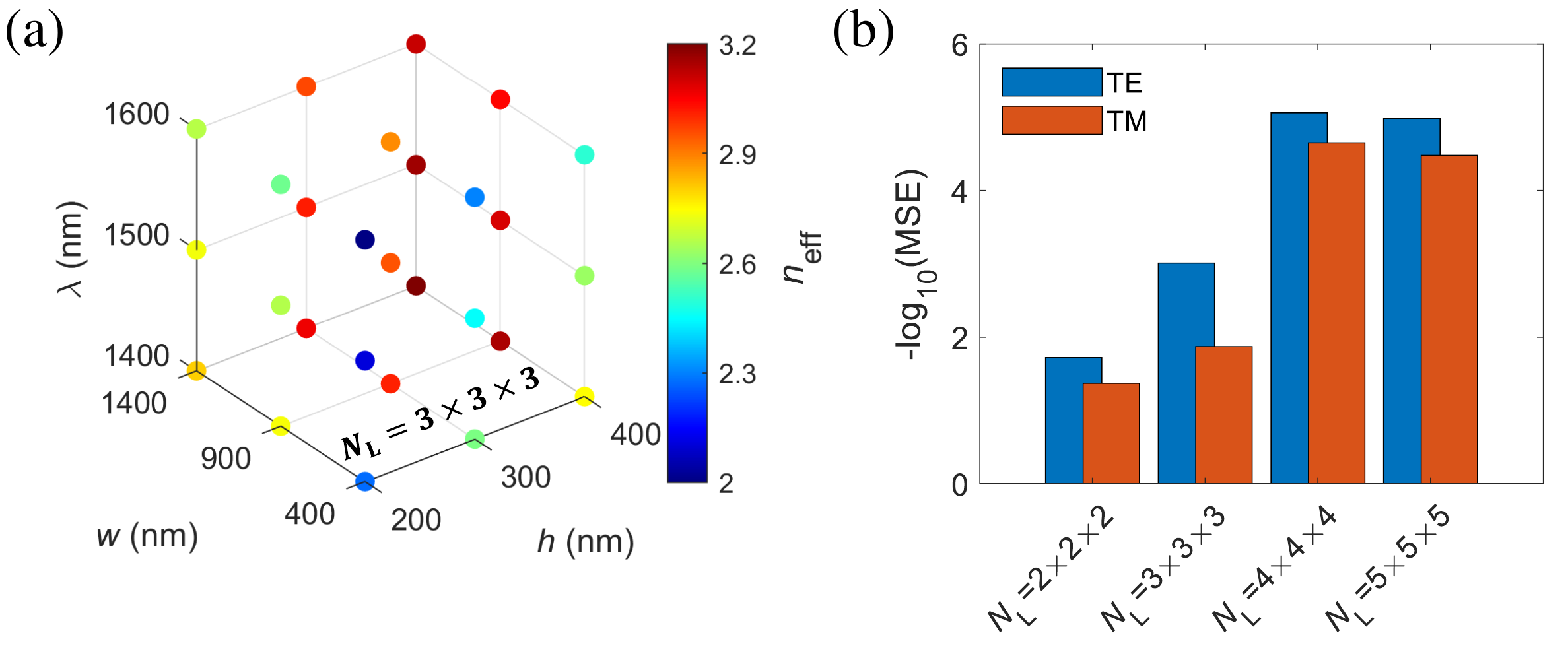}
\caption{\textbf{The mean squared error (MSE) as a function of learning points $N_{\rm{L}}$}. (a) ${N_{\rm{L}}}=3\times 3\times 3$ (total 27) learning points. Note that each circle represents a position as a $f(w,h,\lambda)$. (b) Learning points ${N_{\rm{L}}}=w\times h \times\lambda$, and their corresponding $-{\rm{log}_{10}}$(MSE) when trained with hyperbolic tansigmoid ($tansig$) transfer function with the NN layout as shown in Fig. \ref{fig:NN-network}(c). The blue and orange bars present the MSE for TE and TM polarization, respectively. Other parameters are set to $w=400-1400$~nm, $h=200-400$~nm, and $\lambda=1400-1600$~nm, unless otherwise specified. }
\label{fig:NL-point}
\end{figure}
\section{Neural network (NN) configurations}
Figure~\ref{fig:NN-network} shows the configuration of a neural network and a channel waveguide. We choose silicon (Si) as a channel waveguide (blue) on a silica substrate (grey) with air as cladding, as shown in Fig.~\ref{fig:NN-network}(a). The parameters $w$ and $h$ are the width and height of the waveguide. Note that this configuration is highly compatible with a complementary metal-oxide-semiconductor fabrication process and has a broad range of applications in high-speed optical communication \cite{agrell2016roadmap,marin2017microresonator,jahani2018controlling,mia2019extremely,mia2020exceptional}, bio-sensing \cite{luchansky2012high,lin2013trapping}, light detection and ranging (LIDAR) \cite{sun2013large,kim2018photonic}, and high resolution spectroscopy \cite{yu2018silicon}. However, this model is not limited to Si channel waveguides only; it could be useful in other photonic waveguides. 
Since effective indices of fundamental TE and TM modes define the field/power confinement \cite{jahani2018controlling}, group velocity dispersion \cite{bala2017highly,mia2019extremely}, crosstalks or coupling length \cite{jahani2018controlling,mia2020exceptional}, propagation loss \cite{visser1995modal,shimizu2006fabrication}, phase matching in polarization splitter \cite{sun2016compact,kim2015polarization,zhang2016ultra}, mode conversion in extreme mode converter \cite{kim2018photonic} and free spectral range (FSR) \cite{kim2017dispersion}, to know the $n_{\rm{eff}}$ of TE and TM modes are very significant.

Figure~\ref{fig:NN-network}(b) shows the schematic of the deep learning (DL) model. A DL model is built using neural networks requiring training data (i.e., inputs and outputs) and adequate training, and finally predicts the outputs. In Fig.~\ref{fig:NN-network}(b), we use three inputs, i.e., width $w$, height $h$, and wavelength $\lambda$. After a good training, NN outputs the effective modal indices $n_{\rm{TE}}$ and $n_{\rm{TM}}$ for the TE and TM polarization, respectively. $N_{\rm{L}}$ is the learning point for NN, which comprises of inputs, i.e., ($w$, $h$, $\lambda$) and outputs, i.e., ($n_{\rm{TE}}$, $n_{\rm{TM}}$). The input parameters $n_{\rm{TE}}$ and $n_{\rm{TM}}$ in $N_{\rm{L}}$ are the numerically simulated effective indices using commercially available software \cite{solutions2003lumerical}. 
We use wavelength $\lambda=1400-1600$~nm covering the communication bands, and ($w$, $h$) parameters compatible with the current photolithography and CMOS foundry \cite{jahani2018controlling,mia2020exceptional}. 
Figure~\ref{fig:NN-network}(c) shows the number of layers $N$ and neurons distribution in each layer. Notice that Fig.~\ref{fig:NN-network}(c) is what is hidden in Fig.~\ref{fig:NN-network}(b) with a green rectangular box. The box or NN consists of three hidden layers $N=3$., i.e., Hid. layer $=1,2,3$, and input/output nodes. We consider the number of neurons in the $k^{\rm{th}}$ layer to be $n_{\rm{k}}$, and ${n_{\rm{k}}}=4, 6, 4$ where $k=1, 2$, and $3$, respectively. The blue and green circles present the neurons in each layer and input/output nodes, respectively, as shown in Fig.~\ref{fig:NN-network}(c). The purple circles present the bias (i.e., a constant term), as shown in the same figure. Bias is not coming from a particular neuron and is taken before the learning process, but very useful for the NN. 
A neuron's input in the $k^{\rm{th}}$ layer takes the weighted output from all the neurons in the previous, e.g., ($k-1$) layers with an added bias. 
The relation between the input and output of a neuron is known as the transfer or activation function. We also use hyperbolic tansigmoid ($tansig$) transfer function in hidden layers and input/output nodes. Note that the effects of different transfer functions on the NN will be discussed later in this article. The unknown weights and biases of the NN are obtained after proper training. 
The training is performed using a backpropagation algorithm and adequate matching of input/output with the learning points. To control NN training, we use a learning rate for the networks, which determines how NN modifies a weight by a smaller or larger step size. We use the Levenberg-Marquardt backpropagation algorithm \cite{hagan1994training,Mathworks.com,kicsi2005comparison} for the training purposes and is highly recommended for small-sized feed-forward networks.
\section{Learning points and mean squared error (MSE)}
To know the effective indices as a function of $f(w,h,\lambda)$, is very critical in designing an optics device. For instance, we want to obtain the $n_{\rm{eff}}$ for a Si channel waveguide, as shown in Fig.~\ref{fig:NN-network}(a) as a function of $f(w,h,\lambda)$. 
Our input parameters range are $w=400-1400$~nm, $h=200-400$~nm, and $\lambda=1400-1600$~nm. Thus, we need to perform a numerical simulation \cite{solutions2003lumerical} with $10$ widths (between $400-1400$), $10$ heights (between $200-400$), and $10$ wavelengths (between $1400-1600$) consequently total $1000$ simulations. Then fitting with conventional interpolations (i.e., linear, cubic-spline, and polynomial ) give the effective indices $n_{\rm{eff}}$. However, with the deep learning approach, as shown in Fig.~\ref{fig:NN-network}(b-c), we just need ${N_{\rm{L}}}=3\times 3\times 3$ (total 27) or ${N_{\rm{L}}}=4\times 4\times 4$ (total 64) learning points to predict the $n_{\rm{eff}}$ for the entire input parameters ranges mentioned earlier. The effects of $N_{\rm{L}}$ in predicting $n_{\rm{TE}}$ and $n_{\rm{TM}}$ are shown in Fig~\ref{fig:NL-point}. Figure~\ref{fig:NL-point}(a) presents the distribution of learning points, i.e., ${N_{\rm{L}}}=3\times 3\times 3$ as function of $f(w,h,\lambda)$. For each data along the $w$ axis, there are corresponding three data along the $h$ and three data along the $\lambda$ axis. Figure~\ref{fig:NL-point}(b) mean squared errors (MSE) in log scale as a function of $N_{\rm{L}}$. The blue and orange bars present the MSE for $n_{\rm{TE}}$, and $n_{\rm{TM}}$, respectively. We use the layout, as shown in Fig.~\ref{fig:NN-network}(c) for NN  to obtain the MSE for different $N_{\rm{L}}$. In Fig.~\ref{fig:NL-point}(b), we can see that ${N_{\rm{L}}}=4\times 4\times 4$ has the lowest MSE both for TE and TM polarizations. However, ${N_{\rm{L}}}=5\times 5\times 5$ should provide the lowest errors since it has the maximum learning points. But we choose the layout as given in Fig.~\ref{fig:NN-network}(c). Therefore, different layout configuration and neurons numbers may give the lowest errors when ${N_{\rm{L}}}=5\times 5\times 5$. However, this is not our concern.
More specifically, we want to predict the results with the least amount of learning points. On the contrary, the improvement in MSE for ${N_{\rm{L}}}=3\times 3\times 3$ with the change of layers and neurons is not good enough. Thus, we fix ${N_{\rm{L}}}=4\times 4\times 4$ as the learning points for NN, and all the results afterward follow this $N_{\rm{L}}$. 
The range of input parameters the simulation and NN is shown in Table~\ref{tab:parameter}. 

\begin{table}[htbp]
\centering
\caption{\bf Parameter variation for the waveguide simulation}
\begin{tabular}{ccc}
\hline
Parameter & Variable & Range (nm) \\
\hline
Si waveguide width & $\it w$ & $400 - 1400$ \\
Si waveguide height & $\it h$ & $200 - 400$ \\
Operating wavelength  & $\it \lambda$ & $1400 - 1600$ \\
\hline
\end{tabular}
  \label{tab:parameter}
\end{table}
\begin{figure}[ht!]
\centering
\includegraphics[width=\linewidth]{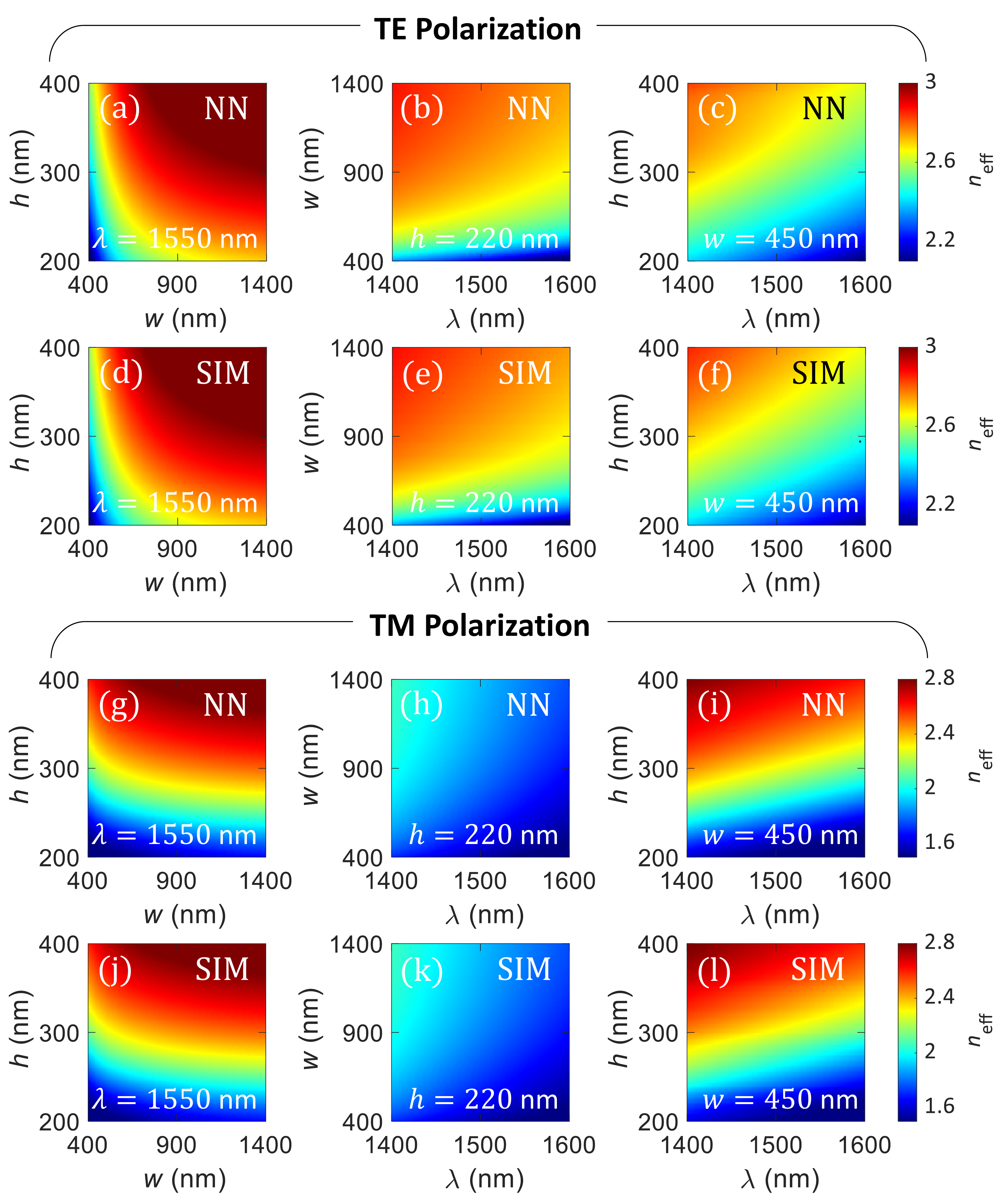}
\caption{\textbf{Performance of NN to predict the effective mode indices $n_{\rm{eff}}$}. (a-l) $n_{\rm{eff}}$ for (a-f) TE polarizations, and (g-l) TM polarizations: (a-c,g-i) prediction from the NN, and (d-f,j-l) their corresponding results from numerical simulation. (a,d,g,j) $w=400-1400$~nm, $h=200-400$~nm while fixing $\lambda=1550$~nm, (b,e,h,k) $\lambda=1400-1600$~nm, $w=400-1400$~nm while fixing $h=220$~nm, and (c,f,i,l) $\lambda=1400-1600$~nm, $h=200-400$~nm while fixing $w=450$~nm. Other parameters are set to ${N_{\rm{L}}}=4\times 4\times 4$ and layout$=4\times 6\times 4$, unless otherwise specified.
}
\label{fig:TEM-neff}
\end{figure}
\begin{figure}[ht!]
\centering
\includegraphics[width=\linewidth]{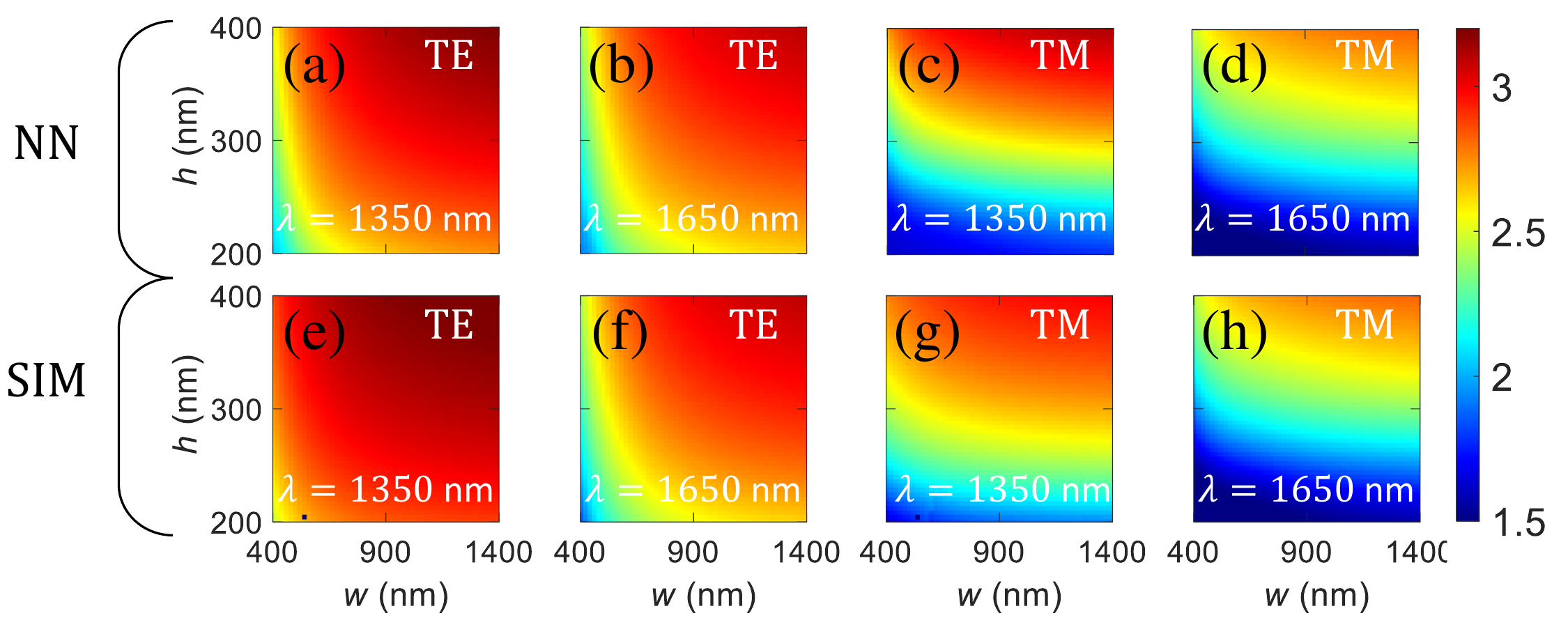}
\caption{\textbf{NN performance outside the training data range}. 
$n_{\rm{eff}}$ from the (a-d) NN model (e-h) full wave simulation: (a,b,e,f) for TE modes, and (c,d,g,h) for TM modes.
Note that, in (a,c,e,g), we fix $\lambda=1350$~nm while in (b,d,f,h) we fix, $\lambda=1650$~nm.  
Other parameters are same as in Fig.~\ref{fig:TEM-neff}, unless otherwise specified.
}
\label{fig:TEM-neff2}
\end{figure}
\section{Performances of NN}
Figure~\ref{fig:TEM-neff} shows the performance of NN in predicting the effective indices $n_{\rm{eff}}$ both for TE and TM polarizations. Note that we use ${N_{\rm{L}}}=4\times 4\times 4$, $tansig$ transfer function, and $4\times6\times4$ as NN layout for these data predictions. Figures~\ref{fig:TEM-neff}(a-f) show the $n_{\rm{eff}}$ for TE polarization, where (a-c) are from the NN predictions and (d-f) present their corresponding simulation results with geometric parameters: (a,d) $w=400-1400$~nm and $h=200-400$~nm while fixing $\lambda=1550$~nm, (b,e) $w=400-1400$~nm and $\lambda=1400-1600$~nm while fixing $h=220$~nm, and (c,f) $\lambda=1400-1600$~nm and $h=200-400$~nm while fixing $w=450$~nm. Figures~\ref{fig:TEM-neff}(g-l) present the $n_{\rm{eff}}$ for the TM polarization while using the same parameters as in Figs.~\ref{fig:TEM-neff}(a-f), respectively. Notice that visually there is no discrepancy between the simulated results and predictions from the NN. Though, we fix the parameters as $\lambda=1550$~nm (Figs.~\ref{fig:TEM-neff}(a,d,g,j)), $h=220$~nm (Figs.~\ref{fig:TEM-neff}(b,e,h,j)), and $w=450$~nm (Figs.~\ref{fig:TEM-neff}(c,f,i,l)), but in practical, we can use any number within the input parameters range. Moreover, the input parameters are equally spaced. To illustrate this, let us consider for an example, ${N_{\rm{L}}}=3\times 3\times 3$, as shown in Fig.~\ref{fig:NL-point}(a). Here, $w$ is spaced equally, i.e., 400, 900, and 1400~nm, so as the $h$, and $\lambda$ axis, as shown in the same figure. So, when we predict the data, we use a different number, e.g., $w=450$~nm, $h=220$~nm, and $\lambda=1550$~nm for the proper justification of NN predictions. Hence, NN is very stable to predict the exact results for $n_{\rm{eff}}$. The degree of accuracy increases with $N_{\rm{L}}$ increments. But at the same time, we need to consider the number of layers (hidden layers) $N$ and the neurons in each layer. 
The MSE for the training data range is $\approx10^{-6}$ and is discussed in section 7.
Note that NN prediction takes $\rm{ms}$ to execute while Lumerical Mode Solutions (with proper mesh) takes $\approx10$~mins (20 points) on a PC with Intel i7 3 GHz processor and 16 Gb of RAM.

To understand the robustness of our model, we explore the NN model with different parameters outside the trained region, i.e., $w<400$ \& $>1400$~nm, $h<200$ \& $>400$~nm, and $\lambda<1400$ \& $>1600$~nm.
For example, we use $\lambda=1350$~nm and $1650$~nm, and these are selected, so they are not a part of training data sets, but random. Figure~\ref{fig:TEM-neff2} shows the results of $n_{\rm{eff}}$ for TE and TM modes when $\lambda=1350$~nm and $1650$~nm.
Figures~\ref{fig:TEM-neff2}(a-d) show the NN predictions and the corresponding simulation results are shown in Figs.~\ref{fig:TEM-neff2}(e-h): (a,b,e,f) are for TE modes and (c,d,g,h) are for TM modes. 
Generally, there is a good agreement between the NN prediction and simulations. 
Note that MSE for $\lambda=1350$~nm is $4.5\times10^{-4}$ (TE) and $5\times10^{-3}$ (TM) whereas $3.62\times10^{-4}$ (TE) and $4.6\times10^{-3}$ (TM) when $\lambda=1650$~nm.
We can also use $w$ and $h$ outside the training range to obtain the results. 
However, too much reduction in $w$ and $h$ may increase MSE, since these parameters are more sensitive than $\lambda$.
Therefore, the NN model is guaranteed within the input parameters
range and can also be used outside the training region but with
caution.
\begin{figure}[ht]
\centering
\includegraphics[width=1\linewidth]{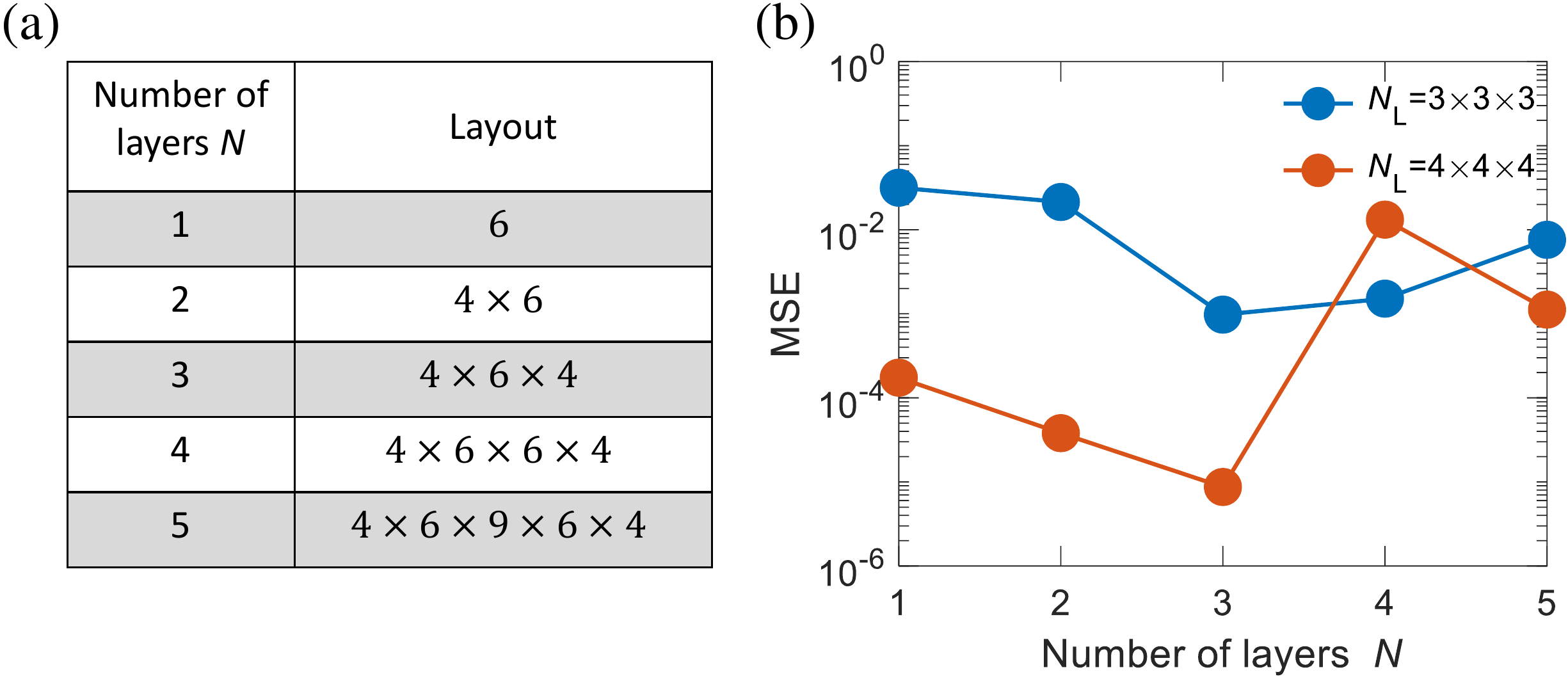}
\caption{\textbf{Best layout for NN (layout with the least MSE).} (a)The number of layers $N$ and their corresponding layout with the different number of neurons in each layer. (b) MSE as a function of $N$. We choose ${N_{\rm{L}}}=3\times 3\times 3$ (dots with solid blue) and ${N_{\rm{L}}}=4\times 4\times 4$ (dots with solid orange) when trained with $tansig$ function. The best layout for NN in (a) corresponding to the least MSE in (b).
}
\label{fig:Best-layout}
\end{figure}
\begin{figure}[ht]
\centering
\includegraphics[width=0.75\linewidth]{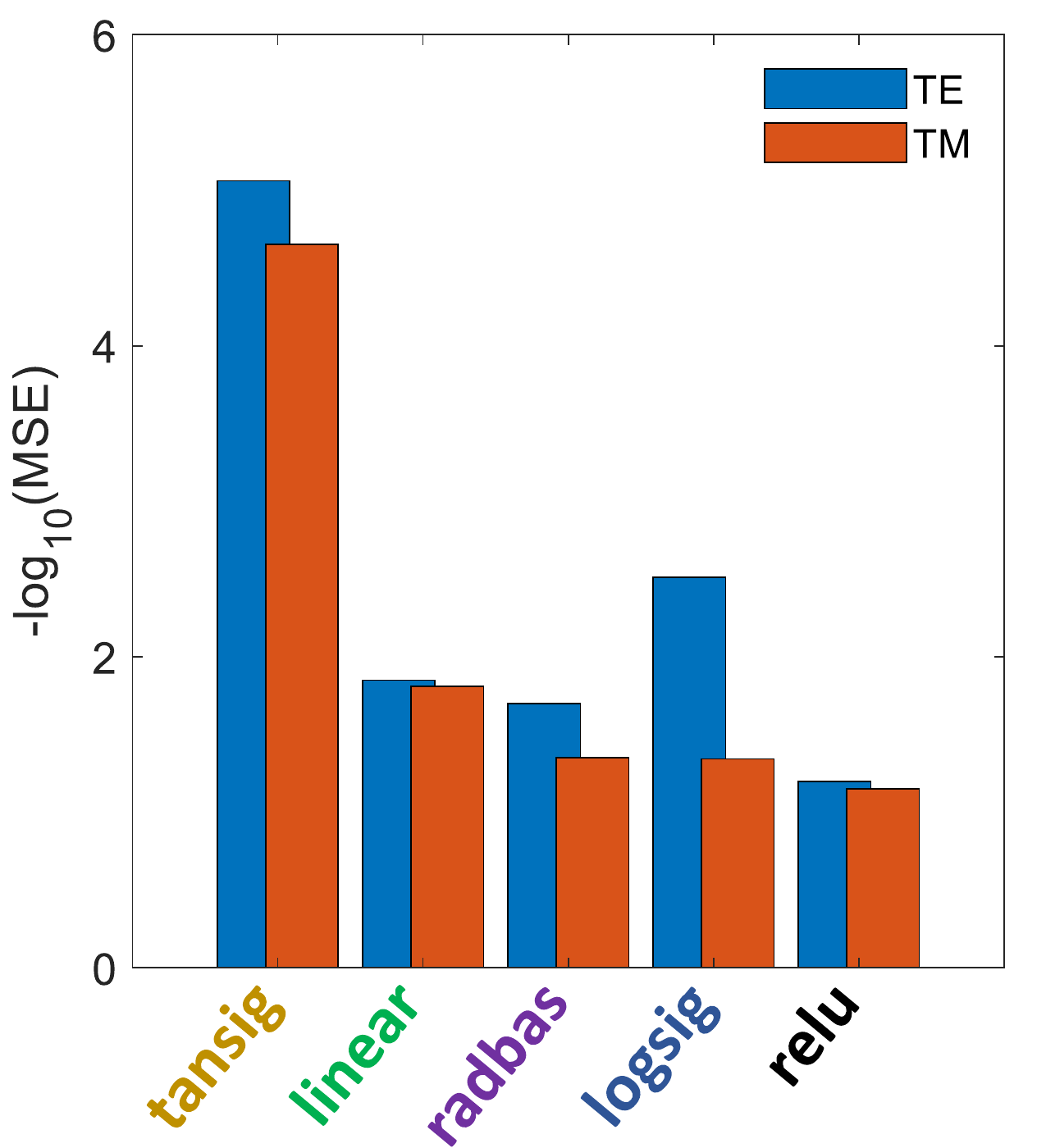}
\caption{\textbf{Transfer functions and MSE.} NN is trained with different transfer functions (labeled in different colors), i.e, $tansig$, $linear$, $radbas$, $logsig$, and $relu$ and their corresponding MSE in log scale. The blue and orange bars present the TE and TM polarization, respectively.
}
\label{fig:Transfer-fun}
\end{figure}
\section{Best possible NN layout}
We choose the NN layout as $4\times 6\times 4$, and Fig.~\ref{fig:Best-layout} shows the reason for our choice. But, there is no generic way to determine the best layout for NN, i.e., number of neurons and layers. Even there is no appropriate guidance to choose a starting point. However, there are some methods (i.e., rule-of-thumb) to help the optimizations. We consider the number of neurons in any layers to be $n_i$, and $n_i$ is chosen as the following \cite{cybenko1989approximation,hornik1989multilayer,hinton2006fast,heaton2008introduction}:
\begin{equation}
  \begin{cases}
    {\rm{No.}\enspace\rm{input}} \leq n_i  \leq {\rm{No.}\enspace\rm{output}}\\
    n_i\leq \frac{2}{3}({\rm{No.}\enspace\rm{input}}+{\rm{No.}\enspace\rm{output}})\\
    n_i\leq 2\times {\rm{No.}\enspace\rm{input}}\\
    {\rm{No.}\enspace \rm{input}} \leq n_i  \leq 3 \times {\rm{No.}\enspace \rm{input}}\\
    
  \end{cases}
\label{eq:thumb}
\end{equation}
For our NN, we have three inputs and two outputs, and we choose the rule-of-thumb as ${\rm{No.}\enspace \rm{input}} \leq n_i  \leq 3 \times {\rm{No.}\enspace \rm{input}}$ \cite{heaton2008introduction}. Thus, 3$\leq n_i\leq$ 9, i.e., 7 possible number of neurons in each hidden layer. The number of possible NN layouts would be $7^N$, where $N$ is the number of layers and would be any positive integer. Figure~\ref{fig:Best-layout}(a) shows the neurons distribution as a function of $N$, and their corresponding MSE is shown in Fig.~\ref{fig:Best-layout}(b). Notice that we use two sets of learning points, i.e., ${N_{\rm{L}}}=3\times 3\times 3$ (dots with solid blue) and ${N_{\rm{L}}}=4\times 4\times 4$ (dots with solid orange), to compare the performance of NN with $N$.  When $N=3$, ${N_{\rm{L}}}=4\times 4\times 4$, the layout $4\times 6 \times 4$ has the least MSE as shown in Fig.~\ref{fig:Best-layout}(b). Thus, it is the best layout, among others as shown in Figs.~\ref{fig:Best-layout}(a-b). More importantly, in NN, when the system has large $N$, it needs a large number of learning points. However, for a smaller network like ours, fewer $N$ could do the job.

\section{Transfer functions}
We use different activation/transfer functions and observe the performance of NN. The used transfer functions are hyperbolic tansigmoid ($tansig$), \textit{linear}, radial basis ($radbas$), log-sigmoid ($logsig$), and rectified linear unit ($relu$). The input and output relations of these transfer functions are as the following ($x:$ input and $y:$ output):
    \begin{align*}
    &tansig:
    y=\frac{2}{1+e^{-2x}}-1\\
    &linear:
    y=x\\
    &radbas:
    y=e^{-x^2}\\
    &logsig:
    y=\frac{1}{1+e^{-x}}\\
    &relu:
    y=x,\enspace x \geq 0\enspace {\rm{and}}\enspace y=0, \enspace x<0
    \end{align*}

\noindent Figure~\ref{fig:Transfer-fun} shows the performance of these transfer functions (labeled in different colors) in predicting the $n_{\rm{eff}}$ while using the layout, as shown in Fig.~\ref{fig:NN-network}(c) and ${N_{\rm{L}}}=4\times 4\times 4$. The blue and orange bars present the TE and TM polarization, respectively. The MSE for the $tansig$ transfer function is the lowest as shown in Fig.~\ref{fig:Transfer-fun}. The MSEs for the other transfer functions follow $logsig<linear<radbas<relu$. Note that $tansig$ outperforms other transfer functions. We use the same transfer function when training the NN at the input/output nodes and the hidden layers. Moreover, $tansig$ is fast enough in predicting effective indices than other transfer functions.
\begin{figure}[ht]
\centering
\includegraphics[width=0.75\linewidth]{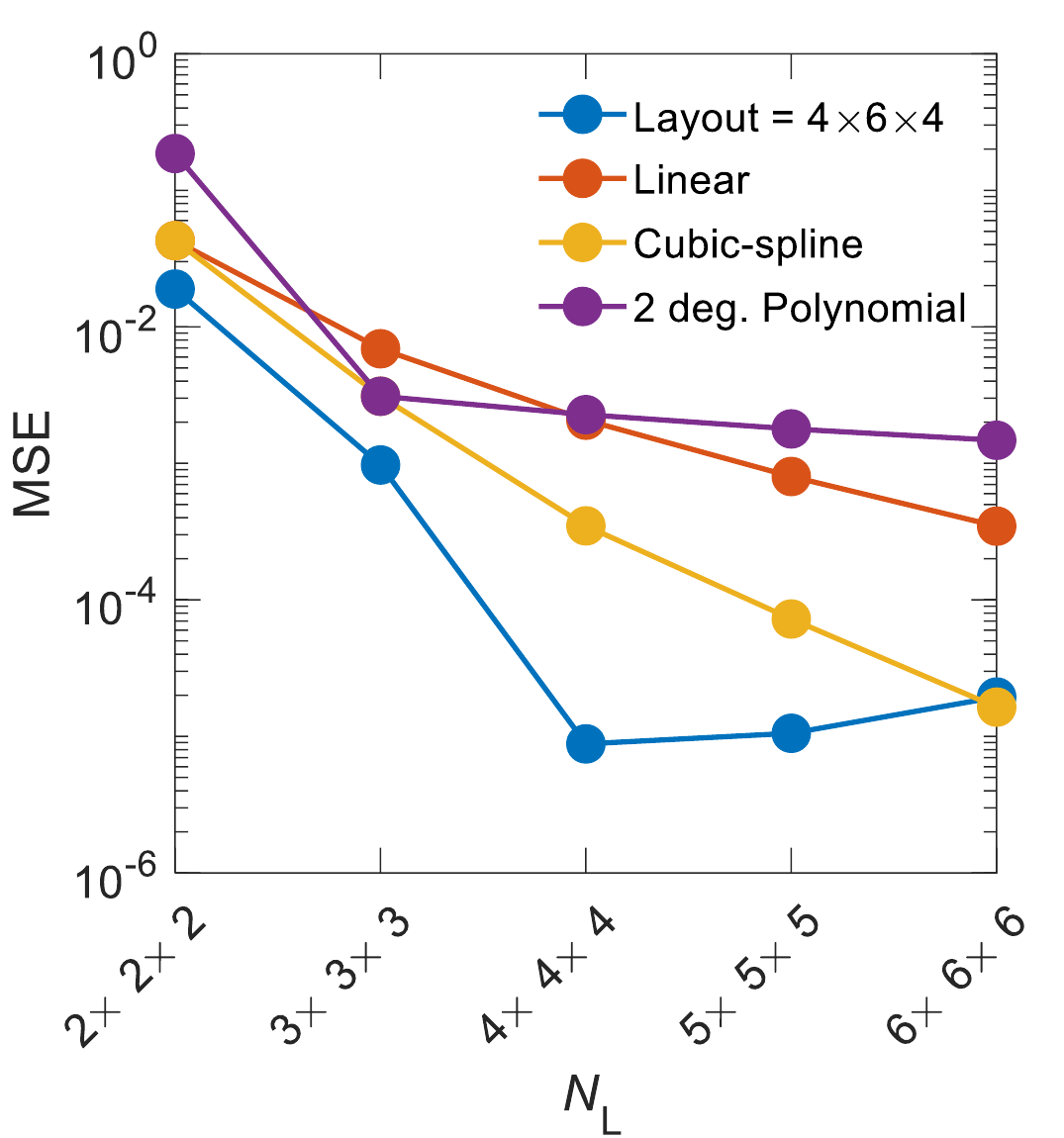}
\caption{\textbf{The proposed NN vs. conventional interpolation techniques.} The MSE for the NN (layout$=4\times 6\times 4$) (dots with solid blue), conventional interpolations: Linear (dots with solid orange), Cubic-spline (dots with solid yellow), and 2 degree Polynomial fitting (dots with solid purple). $N_{\rm{L}}$ presents the learning points for the NN and the corresponding data are made available for the conventional interpolation techniques.
}
\label{fig:ActivMSE}
\end{figure}
\section{Conventional techniques vs. NN}
To compare the NN predictions with the conventional interpolation techniques, we present Fig.~\ref{fig:ActivMSE}. Here, MSE is plotted as a function of learning points $N_{\rm{L}}$. As we can see, linear (dots with solid orange), and 2-degree polynomial fitting (dots with solid purple) have moderate MSE. The cubic-spline (dots with solid yellow) interpolation outperforms linear and polynomial fitting. However, the NN (i.e., layout$=4\times6\times4$) has the lowest MSE (dots with solid blue). Hence, NN outperforms all the conventional interpolation techniques. More importantly, one can modify the NN layout for the $N_{\rm{L}}$ to yield the lowest MSE. Conventional interpolation techniques like linear, polynomial, and cubic-spline try to find the best fits of the learning points. Alternately, a deep learning approach with NN uses the regression model to recognize the pattern and generalize the data sets. The NN training stops when the MSE (e.g., between the desired output and the expected output) is below some threshold value. Moreover, one can suitably choose the layout according to the learning points and optimize it as needed.  
\section{Conclusion}
In conclusion, we present an optical neural network (NN) approach to solve for the fundamental guided mode indices in photonic waveguides (e.g., Si channel) for TE and TM polarizations. With only $4^3$ learning points, we successfully predict the effective indices corresponding to the simulation data points of $10^3$. The mean squared error (MSE) in predicting these data is less than $10^{-5}$. This optical NN can also be applied to predict the indices of higher-order modes and even other parameters. We employ various layouts and show the effect of different transfer functions on the neural network performances. Our a deep learning approach with
the optical neural network could help predict the
results with a wide range of data, but with a few simulations.
\\\\
\noindent\textbf{\large{Disclosures.}} The authors declare no conflicts of interest.
\bibliography{Ref}


\end{document}